\documentstyle[12pt,epsf]{article}

\begin{document}


\begin{titlepage}
\begin{flushright}
\begin{tabular}{l}
FERMILAB--PUB--98/067--T\\
\hspace*{55pt}(extended version)\\
CERN--TH/98--215\\
hep-ph/9802394
\end{tabular}
\end{flushright}
\vskip0.5cm
\begin{center}
  {\Large \bf 
               $B\to\pi$ and $B\to K$ Transitions from\\[5pt]
               QCD Sum Rules on the Light-Cone
  }

\vspace{3cm}
{\sc Patricia~Ball}\footnote{Heisenberg Fellow}
\\[0.3cm]
\vspace*{0.1cm} {\it 
CERN/TH, CH--1211 Gen\`{e}ve 23, Switzerland} \\[2cm]

\vfill

  {\large\bf Abstract:\\[10pt]} \parbox[t]{\textwidth}{ 
I calculate the form factors describing semileptonic and 
penguin-induced decays of $B$ mesons into light pseudoscalar mesons. 
The
  form factors are calculated from QCD sum rules on the light-cone
  including contributions  up to twist~4,
  radiative corrections to the leading twist contribution 
and SU(3)-breaking effects. 
The theoretical uncertainty is estimated
  to be $\sim\,$15\%. The heavy-quark-limit relations between
  semileptonic and penguin form factors are found to be valid in
  the full accessible range of momentum transfer.
}

\end{center}
\end{titlepage}

\noindent{\large\bf 1.}
Decays of $B$ mesons into light mesons offer
the possibility to access the less well known entries in the CKM quark
mixing matrix such as $V_{ub}$ and $V_{ts}$. The measurement of rare
penguin-induced $B$ decays may also give hints at new physics in
the form of loop-induced effects. With new data of hitherto unknown
precision from the new experimental facilities
BaBar at SLAC and Belle at KEK expected to be available in the near
future, the demands for accuracy of theoretical predictions are
ever increasing. The central problem of all such
predictions, our failure to solve non-perturbative QCD,
is well known and so far prevents a rigorous calculation of form
factors from first principles. Theorists thus
concentrate on providing various approximations. The
maybe most prominent of these, simulations of QCD on the lattice, 
have experienced considerable
progress over recent years; the current status for $B$ decays is summarized in
\cite{flynn}. It seems, however, unlikely that lattice calculations will
soon overcome their main restriction in describing $b\to u$ and $b\to s$
transitions, namely the effective upper cut-off that  the finite
lattice spacing imposes on the momentum of the final-state meson. The
cut-off restricts lattice predictions of $B$ decay form factors  to rather
large momentum transfer $q^2$ of about 15$\,$GeV$^2$ or larger. The physical
range in $B$ decays, however, extends from 0 to about
20$\,$GeV$^2$, depending on the process; for radiative decays like
$B\to K^*\gamma$ it is exactly 0$\,$GeV$^2$. Still, one may hope to
extract from the lattice data some information on  form factors in
the full physical range, as 
their behaviour at large $q^2$ restricts the shape
at small $q^2$ via the analytical properties of a properly chosen
vacuum correlation function. The latter function, however, 
also contains  poles and multi-particle cuts whose exact behaviour
is not known, which limits the accuracy of bounds obtained from such
unitarity constraints and until now has restricted their application
to $B\to\pi$ transitions \cite{Laurent,constraints}. The
most optimistic overall theoretical uncertainty one may hope to obtain
from this method is
the one induced by the input lattice results at large $q^2$, which to
date is around (15--20)\% \cite{latticeBpi,Laurent}. A more
model-dependent extension of the lattice
form factors into the low $q^2$ region is discussed in
\cite{latticeparametrizations}.

An alternative approach to heavy-to-light transitions
is offered by QCD sum rules on the
light-cone. In contrast and complementary to lattice simulations, it is
just the fact that the final-state meson {\em does} have large energy
and momentum of order
$\sim m_B/2$ in a large portion of phase-space that is  used as
starting point (which restricts the method to not too large momentum
transfer, to be quantified below). 
The key-idea is to consider $b\to u$ and $b\to s$
transitions as hard exclusive QCD processes and to combine the
well-developed description of such processes in terms of perturbative
amplitudes and non-perturbative hadronic distribution amplitudes
\cite{exclusive} (see also \cite{sterman} for a nice
introduction) with the method of QCD sum rules \cite{SVZ} to
describe the decaying hadron. The idea of such ``light-cone sum
rules'' was first
formulated and carried out in \cite{BBK} in a different context 
for the process $\Sigma\to
p\gamma$, and its first application to $B$ decays was given in
\cite{chernB}. Subsequently, light-cone sum rules were considered for
many $B$ decay processes, see \cite{VMBreview,KRreview} for
reviews. As light-cone sum rules are based on the light-cone
expansion of a correlation function, they can be systematically
improved by including higher twist contributions and radiative 
corrections to perturbative amplitudes. The first calculation in
\cite{chernB} was done at tree-level and to leading twist~2
accuracy. In \cite{BKR,rest}, twist 3 and 4 contributions were included,
and in \cite{radcorr}, one-loop radiative corrections to the twist~2
contribution to the form factor $f_+^\pi$ were
calculated. In \cite{survey}, the corresponding radiative corrections
to the decays of $B$ mesons into the vector mesons $\rho,K^*,\phi$
were calculated. In \cite{late}, the scalar form factor $f_0^\pi$ was
calculated at tree-level. In the
present letter, I calculate the remaining radiative corrections to 
all semileptonic and penguin
$B\to\pi$ and $B\to K$ transitions  and present new and more accurate
results for the corresponding form factors.

\bigskip

\noindent{\large\bf 2.} Let me begin by defining the relevant
quantities. The semileptonic form factors are defined as ($q=p_B-p$)
\begin{equation}
\langle P(p) | \bar q \gamma_\mu b | B(p_B)\rangle  =  f_+^P(q^2) \left\{
(p_B+p)_\mu - \frac{m_B^2-m_P^2}{q^2} \, q_\mu \right\} +
\frac{m_B^2-m_P^2}{q^2} \, f_0^P(q^2)\, q_\mu,\label{eq:SL}
\end{equation}
where $P$ stands for the pseudoscalar meson $\pi$ or $K$ and $q=u$ for
the $\pi$ and $q=s$ for the $K$. The penguin form factor is defined as
\begin{eqnarray}
\langle P(p) | \bar q \sigma_{\mu\nu} q^\nu (1+\gamma_5) b | B(p_B)\rangle
& \equiv &  \langle P(p) | \bar s \sigma_{\mu\nu} q^\nu b |
B(p_B)\rangle\nonumber\\
& = & i\left\{ (p_B+p)_\mu q^2 - q_\mu (m_B^2-m_K^2)\right\} \,
  \frac{f_T^P(q^2)}{m_B+m_K} \label{eq:fT}.
\end{eqnarray}
The physical range in $q^2$ is $0\leq q^2\leq (m_B-m_P)^2$. Although
there are of course no semileptonic decays $B\to K e \nu$, the above
form factors contribute to, say, $B\to K\ell\bar\ell$.
Recalling the results of perturbative QCD for the $\pi$
electromagnetic form factor as summarized in \cite{sterman}, 
one may suppose that the
dominant contribution to the above form factors be the exchange of a
hard perturbative gluon between, for instance, the $u$ quark and the antiquark;
this possibility was advocated in \cite{szcz}. This is,
however, not the case, and it was pointed out already in 
Ref.~\cite{chernB} that the dominant contribution
comes from the
so-called Feynman mechanism, where the quark created in the weak
decay carries nearly all of the final-state meson's momentum, while
all other quarks are soft, and which bears no perturbative suppression
by factors $\alpha_s/\pi$. In an expansion in the inverse $b$ quark
mass, the contribution from the Feynman mechanism is of the same order
as the gluon-exchange contribution  with momentum fraction
of the quark of order $1-\Lambda_{\rm QCD}/m_b$, but it 
dies off in the strict limit $m_b\to\infty$ due to Sudakov effects.
This means that --- unlike the case of the electromagnetic $\pi$
form factor --- knowledge of the hadron distribution amplitudes
$$
\phi(u,\mu^2)\sim \int_0^{\mu^2}\!\! dk^2_\perp\, \Psi(u,k_\perp),
$$ 
where $\Psi$ is the full Fock-state wave function of the $B$ and
$\pi(K)$, respectively, $u$ is the longitudinal momentum fraction
carried by the ($b$ or $u(s)$) quark, $k_\perp$ is the transverse quark
momentum, is not sufficient to calculate the form factors in the form
of overlap integrals
$$
F \sim \int_0^1 du\,dv\, \phi^*_{\pi(K)}(u)\, T_{\rm hard}(u,v; q^2)\,
\phi_B(v)
$$
(with $T_{\rm hard}\propto 
\alpha_s$).\footnote{Note also that not much is known
about $\phi_B$, whereas the analysis of light meson distribution
amplitudes is facilitated by the fact that it can be organized in an
expansion in conformal spin, much like the partial wave expansion of 
scattering amplitudes in quantum mechanics in rotational spin.}
Instead, in the method of light-cone sum rules, only the light meson
is described by distribution amplitudes.
Logarithms in $k_\perp$ are taken into account by the
evolution of the distribution amplitudes under changes in scale,
powers in $k_\perp$ are taken into account by higher twist
distribution amplitudes. The
$B$ meson, on the other hand, is described, as in QCD sum rules, 
by the pseudoscalar current
$\bar d i\gamma_5 b$ in the unphysical region with virtuality
$p_B^2-m_b^2\sim O(m_b)$, where it can be
treated perturbatively. The real $B$ meson, residing on the physical
cut at $p_B^2=m_B^2$, is then traced by analytical continuation, supplemented
by the standard QCD sum rule tools to enhance its contribution with
respect to that of higher single- or multi-particle states coupling to
the same current.

The starting point for the calculation of the form factors 
in (\ref{eq:SL}) and (\ref{eq:fT})  is thus the 
correlation functions ($j_B = \bar d i\gamma_5 b$):
\begin{eqnarray}
{\rm CF}_V & = & 
i\int d^4y e^{iqy} \langle P(p)|T[\bar q\gamma_\mu b](y) 
j_B^\dagger(0)|0\rangle\ =\ 
\Pi_+^P (q+2p)_\mu + \Pi_-^P q_\mu,\\
{\rm CF}_T & = & 
i\int d^4y e^{iqy} \langle P(p)|T [\bar q\sigma_{\mu\nu} q^\nu b](y)
 j_B^\dagger(x)|0\rangle\ =\ 2 i F_T^P (p_\mu q^2 - (pq) q_\mu),
\end{eqnarray}
which are calculated in an expansion around the light-cone
$x^2=0$. The expansion goes in inverse powers of the $b$ quark
virtuality, which, in order for the light-cone expansion to be
applicable, must be of order $m_b$. This restricts the accessible
range in $q^2$ to $m_b^2-q^2 \stackrel{<}{\sim} O(m_b)$
parametrically. For physical $B$ mesons, I choose $m_b^2-q^2\leq
17\,$GeV$^2$. Note also that for very large $q^2$ the influence of the
next nearby pole ($B^*$ for $f_+^\pi$) becomes more prominent.

It proves convenient to perform the calculation  for an arbitrary weak
vertex $\Gamma=\{\gamma_\mu$, $\sigma_{\mu\nu}q^\nu\}$, 
which, neglecting for the
moment radiative corrections, yields:
\begin{eqnarray}
{\rm CF}_\Gamma & =  & \frac{f_\pi}{4} \int_0^1 \!\! du \left[
-\phi_\pi(u)\vphantom{\frac{f_\pi}{4}} 
Tr(\Gamma S_b(Q) \widehat{p}) \right. \nonumber\\
& &  {}+ \frac{m_\pi^2}{m_u+m_d}
\left\{ -\phi_P(u) Tr(\Gamma S_b(Q))+ \frac{i}{6}\, \phi_\sigma(u)\,
\frac{\partial}{\partial Q_\alpha}
\, Tr(\Gamma S_b(Q)\sigma_{\alpha\beta}) p^\beta\right\}\nonumber\\
& & \left.{} + \left\{
g_1(u) - \int_0^u\!\! dv\, g_2(v)\right\} \frac{\partial^2}{\partial
Q_\alpha \partial Q^\alpha} \, Tr(\Gamma S_b(Q) \widehat{p}) - g_2(u)
\,\frac{\partial}{\partial Q_\alpha} \, Tr(\Gamma S_b(Q)
\gamma_\alpha) \right]\nonumber\\
& & {}+ \frac{f_\pi}{4} \int_0^1\!\! dv \int_0^1 {\cal D}\underline{\alpha}\,
\frac{1}{\widetilde{s}^2} \left[ \frac{4f_{3\pi}}{f_\pi} \, v (pq)
\phi_{3\pi}(\underline{\alpha}) Tr(\Gamma\widehat{p}) + (2
\phi_\perp(\underline{\alpha}) - \phi_\parallel(\underline{\alpha}) )
Tr(\Gamma(\widehat{q} + m_b) \widehat{p})\right.\nonumber\\
& & {} + 2v \left\{
\phi_\parallel(\underline{\alpha}) Tr(\Gamma\widehat{p}\widehat{q}) -
2 (pq) \phi_\perp (\underline{\alpha}) Tr(\Gamma) \right\} + \left\{ 2
\widetilde{\phi}_\perp(\underline{\alpha}) -
\widetilde{\phi}_\parallel(\underline{\alpha}) \right\}
Tr(\Gamma(\widehat{q}+m_b) \widehat{p})\nonumber\\
& & \left. \vphantom{\frac{1}{s^2}} + 4 i v
\widetilde{\phi}_\perp(\underline{\alpha}) Tr(\Gamma
\sigma_{\alpha\beta}) q^\alpha p^\beta\right].\label{eq:LC}
\end{eqnarray}
Explicit expressions for $\Pi_\pm$ and $F_T$ were already obtained in
\cite{BKR,rest}. 
Here $Q=q+\bar u p$, $s = m_b^2-Q^2 = m_b^2-u p_B^2-\bar u q^2$,
${\cal D}\underline{\alpha} = d\alpha_1 d\alpha_2 d\alpha_3
\delta(1-\alpha_1-\alpha_2-\alpha_3)$ and $\widetilde s = m_b^2 - (q +
(\alpha_1+v \alpha_3) p)^2$; $S_b(Q)=(\widehat{Q}+m_b)/(-s)$ is the
$b$ quark propagator. In the above expression, $\phi_{\pi,K}$ is the
leading twist 2 distribution amplitude, $\phi_P$ and $\phi_\sigma$ are
the two-particle distribution amplitudes of twist~3, $g_1$ and $g_2$ those
of twist~4, all of which are defined in \cite{BF1}. 
The twist~3 and 4 two-particle distribution amplitudes are
determined completely in terms of the twist~3 and 4 three-particle
distribution amplitudes $\phi_{3\pi}$, $\phi_{\parallel,\perp}$ and
$\widetilde{\phi}_{\parallel,\perp}$ \cite{BF1}.
Note that in the above expression corrections in the
light meson mass are neglected ($m_\pi^2/(m_u+m_d)$, however, is
expressed in terms of the
quark condensate and taken into account). Their inclusion, of
potential relevance in $B\to K$ transitions, is not
straightforward and requires an extension of the method developed in
Ref.~\cite{BF1} to include meson- and quark-mass corrections in the 
twist~4 distribution
amplitudes. According to \cite{T4}, the 
numerical impact on the form factors is small, around 5\%, and
most pronounced at large $q^2$.

\bigskip

\noindent{\large\bf 3.} It is convenient to calculate also the radiative
corrections for arbitrary weak vertex. To twist~2 accuracy, the light
quarks are massless and carry  only longitudinal momentum. The
one-loop calculation does not occasion any particular technical
complications, but results in bulky expressions  which I
refrain from quoting here. The general structure is, as to be
expected, similar to that for the form factor $f_+^\pi$ obtained in
\cite{radcorr}. 
The separation of perturbative and non-perturbative contributions
introduces an arbitrary logarithmic (infra-red) factorization
scale. The condition that the correlation function be independent of
that scale leads to an evolution
equation for the distribution amplitude, which was first derived and
solved in \cite{exclusive} to leading logarithmic accuracy.
In the present context, with full $O(\alpha_s)$ corrections to the
perturbative 
part included, one has to use the next-to-leading order 
evolution of the distribution
amplitude, which was derived in closed form in \cite{muller}.
A natural choice for the factorization scale is the virtuality of the
$b$ quark,  $\mu_{\rm IR}^2 \sim u\, m_b$. For
technical reasons it is, however, more convenient to choose a fixed
scale like $\mu^2_{\rm IR} = m_B^2-m_b^2$, which is of the same order. 
The numerical impact of changing the scale is
minimal.\footnote{This is in contrast to the $\pi$
electromagnetic form factor, which is rather sensitive to the shape of
the distribution amplitude near the end-points.} The penguin form
factor depends  also on an ultra-violet scale, the
renormalization-scale of the local operator $\bar q
\sigma_{\mu\nu}q^\nu b$ appearing in the effective weak
Lagrangian. A natural choice for this ultra-violet scale is $\mu_{\rm UV} =
m_b$.

As for the radiative corrections, it turns out that they are
dominated by the correction to the pseudoscalar $B$ vertex, which, as
discussed below, yields large cancellations 
against the corresponding corrections to the leptonic $B$ decay
constant $f_B$.

\bigskip

\noindent{\large\bf 4.} Let me now derive the light-cone sum rules. The
correlation functions ${\rm CF}_\Gamma$, calculated for unphysical 
$p_B^2$,
can also be written as dispersion relations over the physical cut. Singling
out the contribution of the $B$ meson, one has for instance for $\Pi_+$:
\begin{equation}
{\rm CF}_{\Pi_+} = \frac{m_B^2f_B}{m_b} \, f_+(q^2) \, \frac{1}{m_B^2-p_B^2}
+ \mbox{\rm higher poles and cuts},
\end{equation}
where $f_B$ is the leptonic decay constant of the $B$ meson,
$f_Bm_B^2=m_b\langle B| j_B^\dagger|0\rangle$. 
In order to
enhance the ground-state $B$ contribution to the right-hand side,
 I perform a Borel-transformation:
\begin{equation}
\widehat{B}\,\frac{1}{s-p_B^2} = \frac{1}{M^2} \exp(-s/M^2),
\end{equation}
with the Borel parameter $M^2$. The next step is to invoke quark-hadron
duality to approximate the contributions of hadrons other than the
ground-state $B$ meson, so that finally
\begin{equation}
\widehat{B}\,{\rm CF}_{\Pi_+} =
\frac{1}{M^2}\,\frac{m_B^2f_B}{m_b}\,f_+(q^2)\,e^{-m_B^2/M^2} +
\frac{1}{M^2}\, \frac{1}{\pi}\int_{s_0}^\infty \!\! ds \, {\rm
Im}{\rm CF}_{\Pi_+}(s) \, \exp(-s/M^2).\label{eq:SR}
\end{equation}
This equation is the light-cone sum rule for $f_+$; those for
$f_0$ and $f_T$ look similar. Here $s_0$ is the so-called continuum
threshold, which separates the ground-state from the continuum
contributions; $s_0$ and $M^2$ are in principle free
parameters of the light-cone sum rules, but they can be fixed by requiring
stability of the sum rule under their change.
 In the present context, one can decrease their influence considerably
by also writing $f_B$ as a QCD sum rule, depending on the same
parameters $s_0$ and $M^2$. From the analysis of the latter sum rule,
one finds $s_0\approx 34\,$GeV$^2$ and $M^2\approx$(4--8)~GeV$^2$.
The 
resulting value for $f_B$ is (150--200)~MeV, in perfect agreement with
the results from lattice simulations. This procedure makes the form
factors largely independent of $m_b$, $s_0$ and $M^2$; the remaining
dependence will be included in the error estimate.
Note also that subtraction of the continuum contribution from both
sides of (\ref{eq:SR}) introduces a lower limit of integration $u\geq
(m_b^2-q^2)/(s_0-q^2)$ in (\ref{eq:LC}), which behaves as
$1-\Lambda_{\rm QCD}/m_b$ for
large $m_b$ and thus corresponds to the dynamical 
configuration of the Feynman mechanism. 

Let me now specify the non-perturbative
input. For the $b$ quark I use the one-loop pole mass $m_b = (4.8\pm
0.1)\,$GeV, which is consistent with a recent determination from
the $\Upsilon$ mesons \cite{kuhn}. For the light mesons, 
 the distribution amplitudes need to be specified. 
Fortunately, conformal symmetry
of massless QCD combined with the nonlocal string operators technique
developed in \cite{string}, provides a very powerful tool to describe
higher twist distribution amplitudes in a mutually consistent and most
economic way (see \cite{BBKT} for a detailed discussion). The
determination of the relevant non-perturbative parameters from QCD sum
rules was pioneered in \cite{CZreport}.
In \cite{BF1}, the twist~3 and 4 $\pi$ distribution amplitudes were
obtained including contributions up to conformal spin 11/2 in terms of 6
independent non-perturbative parameters whose values were determined
from QCD sum rules. The leading twist~2
distribution amplitude, on the other hand, can be expanded in
Gegenbauer polynomials $C^{3/2}_i$:
\begin{equation}
\phi_{\pi,K} = 6u(1-u) \left( 1 + \sum_{i=1}^\infty a_i(\mu)
C^{3/2}_i(2u-1)\right).\label{eq:series}
\end{equation}
The Gegenbauer moments $a_i$ renormalize multiplicatively. For $\pi$,
all odd moments vanish because of the $\pi$'s definite G-parity. 
In practice, one truncates the expansion after the first few terms,
$$
\phi_{\pi,K(n)} = 6u(1-u) \left( 1 + \sum_{i=1}^n a_i(\mu)
C^{3/2}_i(2u-1)\right),
$$
with $n=4$ (for $\pi$) or $n=2$ (for $K$). I will discuss the impact
of this truncation on the form factors later. For now, 
I use the $\pi$ distribution amplitude as obtained in
\cite{BF2} (see also \cite{rady}), 
\begin{equation}
a_2^\pi(1\,{\rm GeV}) = 0.44, \quad a_4^\pi(1\,{\rm GeV}) = 0.25.
\end{equation}
For the $K$, on the other hand, the
non-zero value of the strange quark mass induces non-vanishing values of
the odd moments. I use
\begin{equation}
a_1^K(1\,{\rm GeV}) = 0.17, \quad a_2^K(1\,{\rm GeV}) = 0.2,
\end{equation}
where the first value was obtained in \cite{CZreport} and the second
one comes from an analysis of the
sum rule for the $\pi$ in \cite{BF2}, due account being taken of 
SU(3)-breaking effects.

\begin{figure}
\centerline{\epsffile{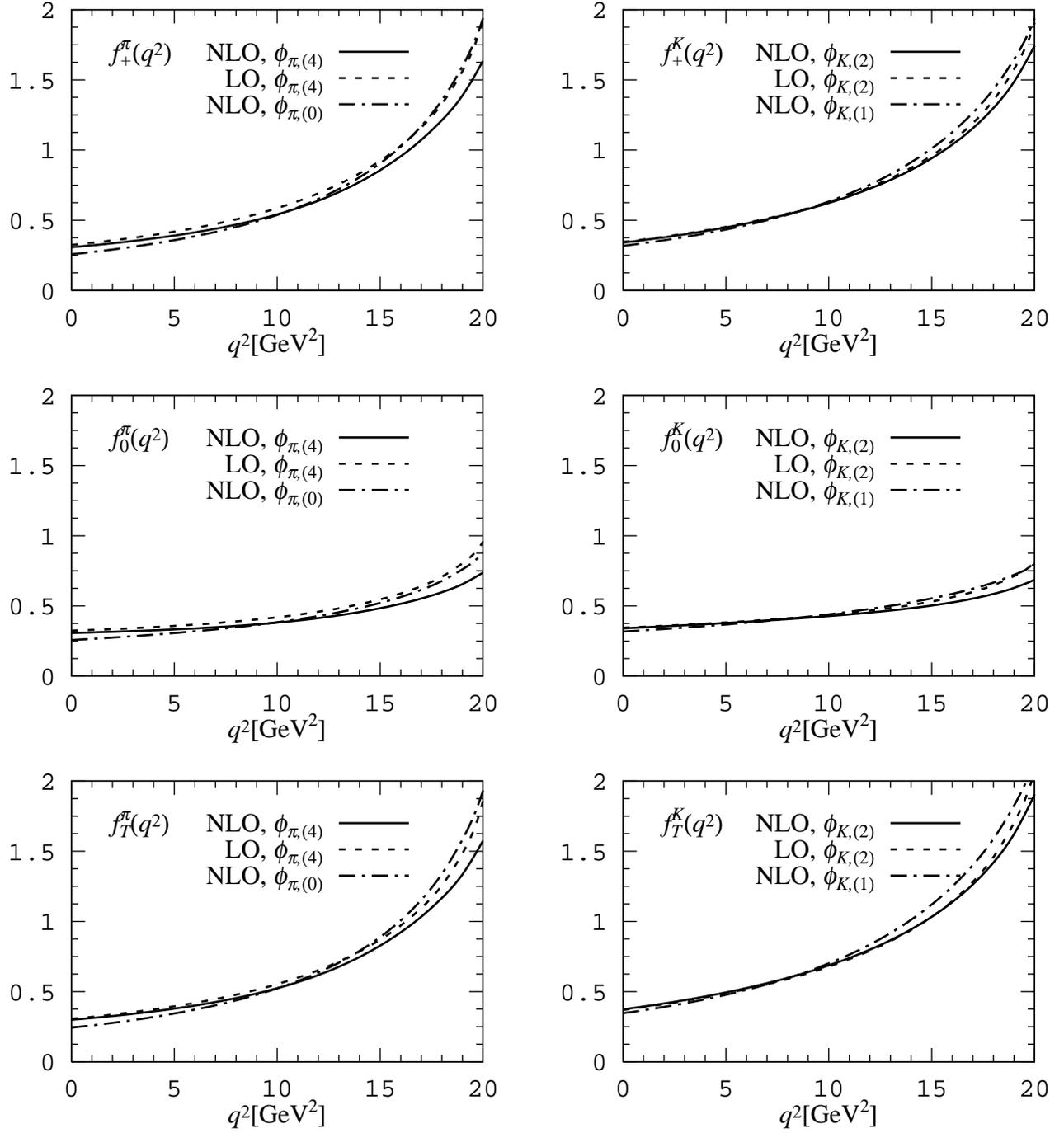}}
\caption[]{Form factors from light-cone sum rules in various approximations.}
\end{figure}

The results are displayed in Fig.~1. The form factor $f_+^\pi$
coincides with the one obtained in \cite{radcorr}. I plot each form
factor using the twist~2 distribution amplitudes as specified above
and with and without $O(\alpha_s)$ corrections, and also using
the asymptotic distribution amplitudes $\phi_{\pi,(0)}$ and
$\phi_{K,(1)}$ to illustrate the impact of non-asymptotic
contributions. 
The plotted curves were obtained with $m_b = 4.8\,$GeV, $s_0
= 33.5\,$GeV$^2$ and $M^2=6\,$GeV$^2$. The distribution amplitudes are
evaluated at the scale $\mu^2 = m_B^2-m_b^2 =: \mu_b^2$. Apparently, the net
effect of radiative corrections on the form factors is rather
small. This is due to an effect already observed in \cite{radcorr}:
the radiative corrections to the QCD sum rule for $f_B$ are rather
large, which is due to the large vertex corrections to the
pseudoscalar $B$ vertex. In the radiative corrections to the
light-cone sum rules, the same vertex appears with corrections of
similar size, so that they cancel between left- and right-hand side of
(\ref{eq:SR}), leaving a net effect of around 10\%. 

For all form factors, the effect of three-particle twist~3 and 4 quark-gluon
contributions (and their induced effects in the two-particle
distribution amplitudes) are small ($\sim 5\% $), so that the
considerable theoretical uncertainty of these terms does not play. This
also shows that the expansion in contributions of increasing twist is
under good control. The remaining twist 3 contributions are
proportional to the quark condensate, which, as already noted in
\cite{radcorr}, introduces only a small uncertainty in the final results.

As is expected from the definition of $f_0$, which refers to
a  scalar current, it increases less sharply in $q^2$ than the other
form factors. A good parametrization for the $q^2$ dependence
 can be given in terms of three parameters as
\begin{equation}\label{eq:para}
F(q^2) = \frac{F(0)}{1-a_F\,(q^2/m_B^2) + b_F \left(
q^2/m_B^2\right)^2}\, .
\end{equation}
\begin{table}
\renewcommand{\arraystretch}{1.4}
\addtolength{\arraycolsep}{3pt}
$$
\begin{array}{|c|cc|cc|cc|cc|}
\hline
& f_+^\pi & f_+^K & f_0^\pi & f_0^K & f_T^\pi(m_b) & f_T^K(m_b) & f_+^\pi\,
\protect{\cite{KRreview}} & f_+^{K,LO}\, \protect{\cite{BKR}}\\ \hline
F(0) & 0.305 & 0.341 & \equiv f_+^\pi(0)  & \equiv
f_+^K(0) & 0.296 & 0.374 & 0.27\pm 0.05 & 0.33\pm 0.05\\
a_F   & 1.29\phantom{0} & 1.41\phantom{0} & \phantom{-}0.266 & 
\phantom{-}0.410 & 1.28\phantom{0} & 1.42\phantom{0} & 1.50 & 1.14\\
b_F   & 0.206 & 0.406 & -0.752 & -0.361 & 0.193 & 0.434 & 0.52 & 0.05\\
\hline
\end{array}
$$
\caption[]{Results for form factors with $m_b=4.8\,$GeV, $s_0 =
  33.5\,$GeV$^2$ and $M^2=6\,$GeV$^2$ in the parametrization of
Eq.~(\protect{\ref{eq:para}}). Renormalization scale for $f_T$ is $\mu
= m_b$. The theoretical uncertainty is $\sim 15\,$\%.}
$$
\begin{array}{|c|cc|cc|}
\hline
q^2 & f_{+,{\rm latt}}^\pi(q^2) \protect{\cite{latticeBpi,Laurent}} & 
f_{+,{\rm LCSR}}^\pi(q^2) & f_{0,{\rm latt}}^\pi(q^2) 
\protect{\cite{latticeBpi,Laurent}} & f_{0,{\rm LCSR}}^\pi(q^2)\\ \hline
14.9\, {\rm GeV}^2 & 0.85\pm0.20 &  0.85\pm 0.15 & 0.46\pm 0.10 & 0.5\pm 0.1\\
17.2\, {\rm GeV}^2 & 1.10\pm 0.27 &  1.1\pm 0.2 & 0.49\pm 0.10 & 0.55\pm 0.15\\
20.0\, {\rm GeV}^2 & 1.72\pm 0.50 &  1.6\phantom{\pm 0.0} 
& 0.56\pm 0.12 & 0.7\phantom{\pm 0.0} \\
\hline
\end{array}
$$
\renewcommand{\arraystretch}{1}
\addtolength{\arraycolsep}{-3pt}
\caption[]{Comparison of lattice results for $B\to\pi$ form factors
with results from light-cone sum rules. The errors for  lattice
results are those quoted in \protect{\cite{Laurent}}.}
\end{table}
The parameters are given in Table~1 for central values of the input
parameters. For
comparison, I also give the results for $f_+^\pi$ quoted in
\cite{KRreview} and $f_+^K$ obtained in \cite{BKR}, the latter being
obtained in leading-logarithmic accuracy. The table confirms what can
also be inferred from the figure, namely that, for both $\pi$ and $K$,
mesons $f_+$ and $f_T$ nearly
coincide. Comparison with the $K$ form factors shows that the main
SU(3)-breaking effect is in the normalization $F(0)$, whereas the
$q^2$ dependence is only slightly modified. This can be understood
from the fact that the formation of a $\pi$ or $K$ meson is proportional to
their respective decay constants $f_{\pi,K}$, so that one would
naively expect an enhancement $\sim f_K/f_\pi=1.2$ of the $K$ form
factors (at least if the three-parton states are not important), which
is essentially what I find.

Varying all input parameters within their respective allowed ranges, I
obtain uncertainties between 5 and 10\%. Combining this with the
systematic uncertainty $\sim$10\% introduced by the need to separate
the ground-state $B$ contribution from that of higher states, the
final uncertainty of the form factors is $\sim$15\%. A slight reduction of
this uncertainty may be possible if more accurate information on the
twist 2 distribution amplitude becomes available, for instance from
lattice simulations.

A comparison with lattice results from the UKQCD collaboration is given in
Table~2. The agreement with the lattice data is
 excellent, as it was also found for $B\to\rho,K^*$ form factors in
 \cite{survey}. The LCSR point at $q^2=20\,$GeV$^2$ is just
for illustration, because of which I also refrain from assigning it
an error.

\bigskip

\begin{figure}
\centerline{\epsffile{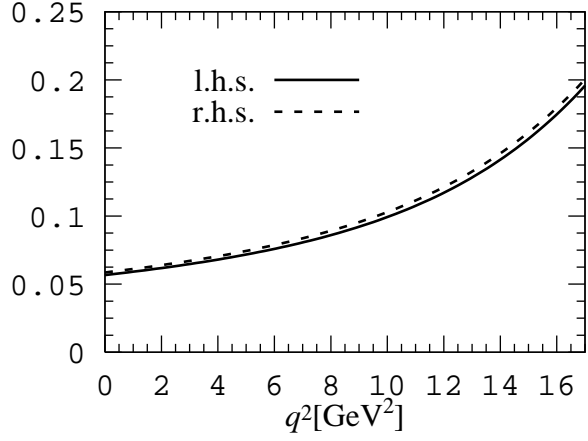}}
\caption[]{Isgur-Wise relation (\protect{\ref{eq:IW}}) between penguin
  and semileptonic form factors.}
\end{figure}

\noindent {\large\bf 5.} In the limit $m_b\to\infty$, Isgur and Wise
have obtained a relation between the semileptonic and the penguin form
factors \cite{IW90}:
\begin{equation}\label{eq:IW}
\frac{f_T}{m_B+m_P} = \frac{1}{2m_b}\left\{
  \left(1+\frac{m_b^2-q^2}{q^2} \right) f_+ -
  \frac{m_b^2-q^2}{q^2}\, f_0 \right\},
\end{equation}
which is strictly valid only
near zero recoil (i.e.\ near $q^2 = m_b^2$). In Fig.~2 I plot the
left- and right-hand sides of Eq.~(\ref{eq:IW}) over the full range of
$q^2$. The agreement between
the curves for {\em all} $q^2$ is striking; they differ by 3\% at
$q^2=0$ and by less than 1\% at $q^2=17\,$GeV$^2$. A closer inspection of
the underlying light-cone sum rules shows that to twist 2 accuracy 
(\ref{eq:IW}) is valid exactly and for arbitrary $m_b$ already at the
level of correlation functions $\Pi_\pm$ and $F_T$ and thus --- to
that accuracy ---  is  independent of the details of the extraction of
the $B$ meson contribution. For the twist 3 and 4 contributions to $\Pi_\pm$
and $F_T$, (\ref{eq:IW}) holds in the kinematical regime
characteristic for the Feynman mechanism, i.e.\ near $u\sim 1$, and up
to terms which are suppressed by one power of $m_b$, which account for
the 3\% breaking of (\ref{eq:IW}) by the light-cone sum rule results. 

\bigskip

\noindent {\large\bf 6.} As the uncertainty associated with the
twist 3 two-particle distribution amplitudes and the higher twist
amplitudes is small, and  the form factor thus depends essentially on
the quality of information on the twist 2 distribution amplitude, it is 
worthwhile to investigate in more detail the impact of
truncating the series (\ref{eq:series}) after the first few terms. Let
me first recall that in \cite{BF2} $a_2^\pi(1\,{\rm GeV})$ was
 determined from a QCD sum rule for Gegenbauer moments of in principle
 arbitrary degree, whereas $a_4^\pi$ was obtained from requiring
$\phi_\pi(1/2,1\,{\rm GeV})=1.2 \pm 0.3$, which follows from a
QCD sum rule for the $\pi NN$-coupling.
The reason for not considering QCD sum rules for higher moments is
that they show a strong divergence with the degree $n$, rendering the
series in (\ref{eq:series}) highly divergent. In order to simulate
possible effects of higher moments without distorting the asymptotic
$u(1-u)$ behaviour near the end-points too much, I allow for a
logarithmic divergence of the sum in Gegenbauer polynomials, yielding
 the following models:
\begin{eqnarray}
{\rm model\ I:} & & \phi_\pi(1/2,\mu=1\,{\rm GeV})\ =\ 1.2,\ a_2^\pi 
\mbox{\rm\  model-dependent:}\nonumber\\
\phi^{I}_\pi(u,\mu=\mu_b) & = & 6u(1-u) - 0.95\cdot 6u(1-u)\,
\frac{3}{5}\,\ln u \ln (1-u),\\
{\rm model\ II:} & & \phi_\pi(1/2,\mu=1\,{\rm GeV})\ =\ 1.2,\ a_2(1\,{\rm
  GeV}) = 0.44:\nonumber\\
\phi^{II}_\pi(u,\mu=\mu_b) & = & 6u(1-u) \{1+0.35\,
C^{3/2}_2(2u-1)\}\nonumber\\
& & {} - 8.5\cdot 6u(1-u)\, \left\{ \frac{3}{5}\,\ln u \ln (1-u) + 1 +
  \frac{7}{50}\, C^{3/2}_2(2u-1)\right\}.
\end{eqnarray}
The corresponding Gegenbauer-spectra fall off as $1/n^3$:
\begin{eqnarray*}
{\rm (I)}: \: \{a_n\} & = & \{ 1,0.13,0.030,0.011,\dots \}\\
{\rm (II)}: \: \{a_n\} & = & \{ 1,0.35,0.27,0.10,0.049,0.027,\dots \}
\end{eqnarray*}
\begin{figure}
\centerline{\epsffile{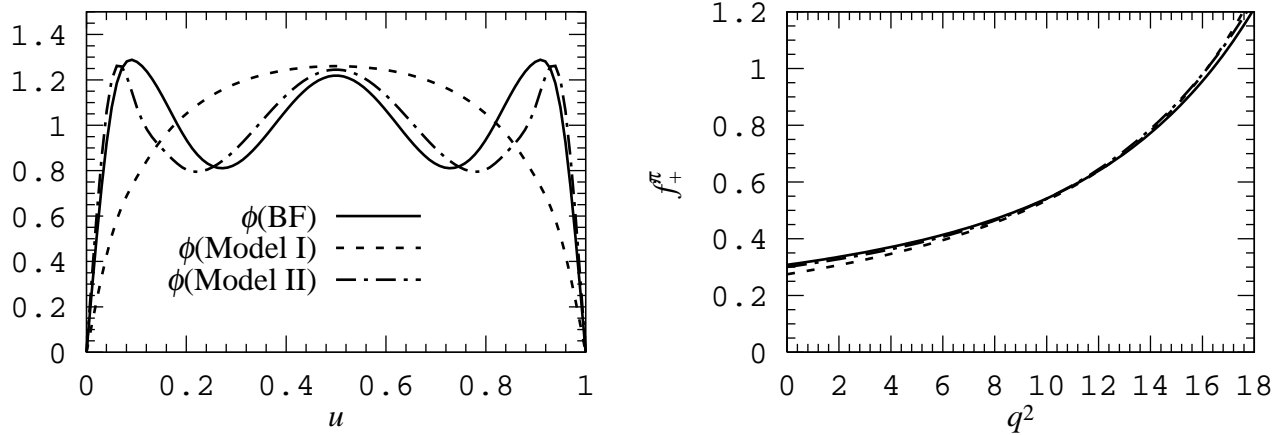}}
\caption[]{Dependence of $f_+^\pi$ on the twist 2 $\pi$ distribution 
amplitude.}
\end{figure}
In Fig.~3 I plot the above model distribution amplitudes as well as
the one suggested by Braun and Filyanov \cite{BF2}, i.e.\
$\phi_{\pi(4)}$. Although they look rather different, the resulting
form factors $f_+^\pi$, also shown in Fig.~3, vary by at most 10\%,
i.e.\ are within the theoretical error. 
It is  evident that the form factors do not depend on the details of
the Gegenbauer-spectrum, but are sensitive only to a few gross
characteristics like the value of the distribution amplitude at one
point (different from the end-points) and the first one or two moments.
This is true as long as the distribution amplitudes are folded with
smooth functions (as it is the case for form factors), so that higher
order oscillatory Gegenbauer polynomials are effectively ``washed out''. 
A determination of the relevant few characteristics from an independent source,
e.g.\ lattice simulations, would evidently help to further increase
the accuracy of form factors calculated from light-cone sum rules.

\bigskip

\noindent {\large\bf 7.} Summarizing, I have calculated the
semileptonic and penguin form factors of $B\to\pi$ and $B\to K$
transitions from light-cone sum rules. A new feature was the inclusion
of one-loop radiative corrections to the leading twist
contributions. The results are summarized in Fig.~1 and Table~1. The
impact of radiative corrections and higher twist contributions is
small, so that the achievable accuracy is limited by the inherent
systematic uncertainty of light-cone sum rules, which is associated
with the extraction of the $B$ meson ground-state contribution out of
the continuum of states coupling to the same current. This uncertainty
is estimated to be $\sim\,$10\% and of the same size as the
uncertainty induced by the input parameters in the sum rule. Hence,
further refinement of the calculation by including higher twist contributions
or two-loop radiative corrections is not expected to yield higher 
accuracy of the result. It would, however, be useful to have an
independent determination of the few lowest moments of the twist~2
$\pi$ and $K$ meson
distribution amplitudes from lattice simulations. The existing results
\cite{latticeDA} have large uncertainties, and in view of the recent
improvements of the methods of lattice QCD
and the availability of much more powerful
computers, more accurate results seem within reach.
 Very recently \cite{guido}, a new method was suggested 
to calculate the leading twist distribution amplitude on the lattice
directly as a function of $u$. If feasible with statistical and
systematic errors in the 20\% range, this would  help to
reduce the total uncertainty of the $B\to\pi,K$ form factors to $\sim\,$10\%.
 The application of these lattice results
would not be restricted to $B$
meson decays, but also of direct relevance to the description of
other hard exclusive
processes, for instance single-meson production at HERA.

\bigskip

\noindent{\bf Acknowledgements:} I thank V.M.\ Braun for discussions and
NORDITA, Copenhagen, Denmark, for financial support
during my stay there. Fermilab is
operated by Universities Research Association,
Inc., under contract no.\ DE--AC02--76CH03000
with the U.S.\ Department of Energy.

\end{document}